\documentclass[english,twocolumn,aps,prb]{revtex4}
\usepackage[T1]{fontenc}
\usepackage[latin9]{inputenc}
\usepackage{graphicx}
\usepackage{amssymb}

\makeatletter

\makeatletter



\makeatother

\usepackage{babel}
\makeatother

\begin{document}

\title{Eu$^{2+}$ spin dynamics in the filled skutterudites
EuM$_{4}$Sb$_{12}$ (M$=$Fe, Ru, Os)}

\author{F. A. Garcia$^{1}$, C. Adriano$^{1}$, G. G. Cabrera$^{1}$, L.
M. Holanda$^{1}$, P. G. Pagliuso$^{1}$, M. A. Avila$^{2}$, S. B.
Oseroff$^{3}$, and C. Rettori$^{1,2}$}

\affiliation{$^{1}$Instituto de Física {}``Gleb Wataghin'', UNICAMP, Campinas-SP,
13083-970, Brazil.\\
 $^{2}$Centro de Ciências Naturais e Humanas, Universidade Federal
do ABC, Santo André-SP 09210-170, Brazil.\\
 $^{3}$Department of Physics, San Diego State University, San Diego,
California 92182, USA.}

\begin{abstract}
We report evidence for a close relation between the thermal activation
of the rattling motion of the filler guest atoms, and inhomogeneous
spin dynamics of the Eu$^{2+}$ spins. The spin dynamics is probed
directly by means of Eu$^{2+}$ electron spin resonance (ESR), performed
in both $X$-band ($\approx9.4$ GHz) and $Q$-band ($\approx34$
GHz) frequencies in the temperature interval $4.2\lesssim T\lesssim300$
K. A comparative study with ESR measurements on the $\beta$-Eu$_{8}$Ga$_{16}$Ge$_{30}$
clathrate compound is presented. Our results point to a correlation
between the rattling motion and the spin dynamics which may be relevant
for the general understanding of the dynamics of cage systems. 
\end{abstract}
\maketitle

\section{Introduction}

The discovery of localized phonon modes in metallic compounds opens
an avenue for the study of electron-phonon and phonon-phonon interactions
in solids. The family of the filled skutterudite compounds are among
the materials where this scenario is believed to take place \citep{Keppens}.
These are cage systems inside which a guest, or filler, atom may perform
relative large excursions. These excursions are described in terms
of localized and isolated phonon modes (Einstein modes), usually called
rattling modes. In this sense, the guest is referred as the rattler
atom and its dynamics are fully characterized by a single parameter
$\theta_{E}$, the Einstein temperature \citep{Sales}.

Filled skutterudite compounds have the general formula RT$_{4}$X$_{12}$,
where usually R is a rare earth or acnitide; T is a transition metal
(Fe, Ru, Os or Co), and X is a pnictogen (P, As or Sb). These compounds
crystallize in the LaFe$_{4}$P$_{12}$ structure with space group
$Im3$ and local point symmetry T$_{h}$ for the R ions \citep{Jeitschko}.
This structure hosts a wide range of physical properties, including
exotic strongly correlated ground states \citep{Maple} as well as
presenting a promising potential for application in the construction
of thermoelectric devices \citep{Sales}.

The latter feature is argued to be closely related to the dynamics
of the rattler R atom \citep{Snyder}. The rattling modes seem to
promote the dampening of the thermal conductivity mainly through the
incoherent scattering of the Debye phonons. This leads to the concept
of a \emph{phonon-glass} type of heat conduction, which gives a fair
picture of experimental results. Two important approximations to the
rattler ion excursions are commonly adopted in this scenario: a description
in terms of isolated (non correlated) and localized (non dispersive)
phonon modes \citep{Keppens}.

Extensions of this scenario have been a matter of intense discussion
in the field. One may find strong experimental evidence for the dispersion
of these modes \citep{Koza,Schenelle1, Schenelle2} which raises the
question of the validity of the independent rattler approximation.
It also raises the question of weather or not an {}``electron glass'',
meaning a coupling between the guest ion dynamics and electronic degrees
of freedom, could also be realized in these materials. Indeed, there
are some recent theoretical proposals in this direction \citep{Hotta, Ueda, Kondo}.
In experiments, there is evidence for an interplay between the rattling
motion and quadrupolar fluctuations with the unconventional superconductivity
in PrOs$_{4}$Sb$_{12}$ \citep{Goto}. Moreover, $^{139}$La nuclear
magnetic resonance (NMR) show a correlation between the $d$ band
of transition metal and the rare earth rattling motion realized through
electron-phonon coupling \citep{Nakai}. In this direction, the rattling
modes may also have an important role in the strongly correlated electronic
phenomena found in cage systems. For applications, a better understanding
of the description of the rattling modes may possibly lead to a more
efficient design of thermoelectric materials.

In a previous work, we presented electron spin resonance (ESR) as
a probe for the guest ion dynamics \citep{GarciaRat}. We showed that
the Yb$^{3+}$ spectra in Ce$_{1-x}$Yb$_{x}$Fe$_{4}$P$_{12}$ ($x\sim0.002$)
clearly reveal the R ion dynamical behavior. Taking advantage of this
novel application of the ESR technique, here, we investigate the EuT$_{4}$Sb$_{12}$
(T = Fe, Ru, Os) skutterudites, to observe whether the Eu$^{2+}$
ESR spectra is a good probe for their rattling behavior. For comparison,
we also present measurements on the $\beta$-Eu$_{8}$Ga$_{16}$Ge$_{30}$
clathrate compound, another well-known cage system inside which Eu$^{2+}$
behaves as a rattler ion.

EuT$_{4}$Sb$_{12}$ (T $=$ Fe, Ru, Os) are metallic systems that
undergo a ferromagnetic transition around $T_{C}=90$ K, $=4$ K and
$=9$ K, respectively \citep{Bauer}. Close investigation of the T
$=$ Fe system indicates that, in fact, a ferrimagnetic transition
takes place at $T_{C}=90$ K, as a consequence of an antiferromagnetic
coupling of the Eu and Fe moments \citep{Krishnamurthy,Krishnamurthy2}.
The rattling behavior of the Eu$^{2+}$ ions was studied by structure
refinement of x-ray diffraction measurements \citep{Bauer} and also
by extended x-ray fine structure (EXAFS) measurements \citep{Cao}.
The results obtained from both methods are in close agreement. The
first measurements gave, for T $=$ Fe, Ru, Os, $\theta_{E}=84$,
$78$, $74$ K respectively, whereas EXAFS gave, for T$=$Ru, Os $\theta_{E}=81$,
$78$ K respectively. In both measurements, no signs of static \emph{off-center}
displacement of the Eu$^{2+}$ were found. X-ray absorption near edge
spectroscopy (XANES) and also susceptibility measurements, point to
a predominant $2+$ valence state of the Eu atoms, although a slightly
mixed valence state cannot be completely excluded ($v\lesssim2.1$)
\citep{Bauer}. In particular, for EuT$_{4}$Sb$_{12}$ (T$=$Ru,
Os), the susceptibility measurements are compatible with the full
Eu$^{2+}$moment in the temperature interval $2\lesssim T\lesssim400$K.

The type-I clathrate compound $\beta$-Eu$_{8}$Ga$_{16}$Ge$_{30}$
is a metallic compound which undergoes a ferromagnetic transition
at $T_{c}=35$ K. In type-I clathrates, one finds two distinct types
of cages that eventually lead to different dynamical behaviors of
the guest elements. In particular, in $\beta$-Eu$_{8}$Ga$_{16}$Ge$_{30}$,
the majority of the guest Eu$^{2+}$ ions are subjected to a potential
with four symmetric \emph{off-center} energy minima. In these cages,
it was found that the dynamical behavior of the Eu$^{2+}$ ions includes
not only an \emph{off-center} thermal activated rattling mode, but
also quantum tunneling between the low temperature potential energy
minima \citep{Sales2,Baumbach}.

In this work, we report a non-trivial evolution of the X-band ESR
spectra of EuT$_{4}$Sb$_{12}$ (T $=$ Ru, Os). We shall discuss
that, in the low temperature region ($4.2\lesssim T\lesssim150$ K),
the rattling motion couples with the Eu$^{2+}$ spins. This gives
rise to inhomogeneous spin dynamics, which manifests in the ESR linewidth
($\Delta H$) behavior. We shall also discuss how our Q-band ESR spectra
give support to these findings.

\section{Experiment}

For this work, we use single crystals of EuT$_{4}$Sb$_{12}$ (T$=$Fe,
Ru, Os) grown in Sb-flux as described in Ref. \citep{Bauer}. The
resulting filled skutterudite structure was checked by x-ray powder
diffraction. High filling rates were confirmed by refinement methods.
The ESR experiments were carried on crushed small pieces of single
crystals with high grain size homogeneity. The ESR spectra were taken
in a X-band and Q-band Bruker spectrometer using appropriate resonators
coupled to a $T$-controller of helium gas flux system. The experiments
covered the temperature interval $4.2\lesssim T\lesssim300$ K. For
comparison, X-band ESR experiments were performed on a single crystal
of the clathrate compound $\beta$-Eu$_{8}$Ga$_{16}$Ge$_{30}$.

ESR detects the power P absorbed from the transverse magnetic microwave
field as a function of the static magnetic field $H$. The sensitive
of the instrument is improved by applying a lock-in technique with
field modulation. As a result, it is the absorption derivative $\frac{dP}{dH}$
which is observed . In our experiments, the ESR spectra showed a single
Dysonian lineshape, described by equation:

\begin{equation}
P(H)\propto\frac{\Delta H+\alpha(H-H_{res})}{(H-H_{res})^{2}+\Delta H^{2}}\label{eq: dyson}\end{equation}

this lineshape contains an $\alpha=D/A$ parameter expressing the
ratio between the dispersion ($D$) and absorption ($A$) of the microwave
radiation when it probes a metallic material \citep{Barnes}. This
$D/A$ parameter appears due to skin depth effects. A small excess
of EuRu$_{4}$Sb$_{12}$ was synthesized and the resulting crystals
were crushed and sieved into fine powder. The powdered crystals were
investigated in X-band and we obtained very similar results, but in
this case the lineshape was completely symmetric ($D=0,$ no skin
depth effects). This investigation was important rule out any surface
effect in our results.

\section{Results And Discussion}

\begin{figure}
\begin{centering}
\includegraphics[scale=0.22]{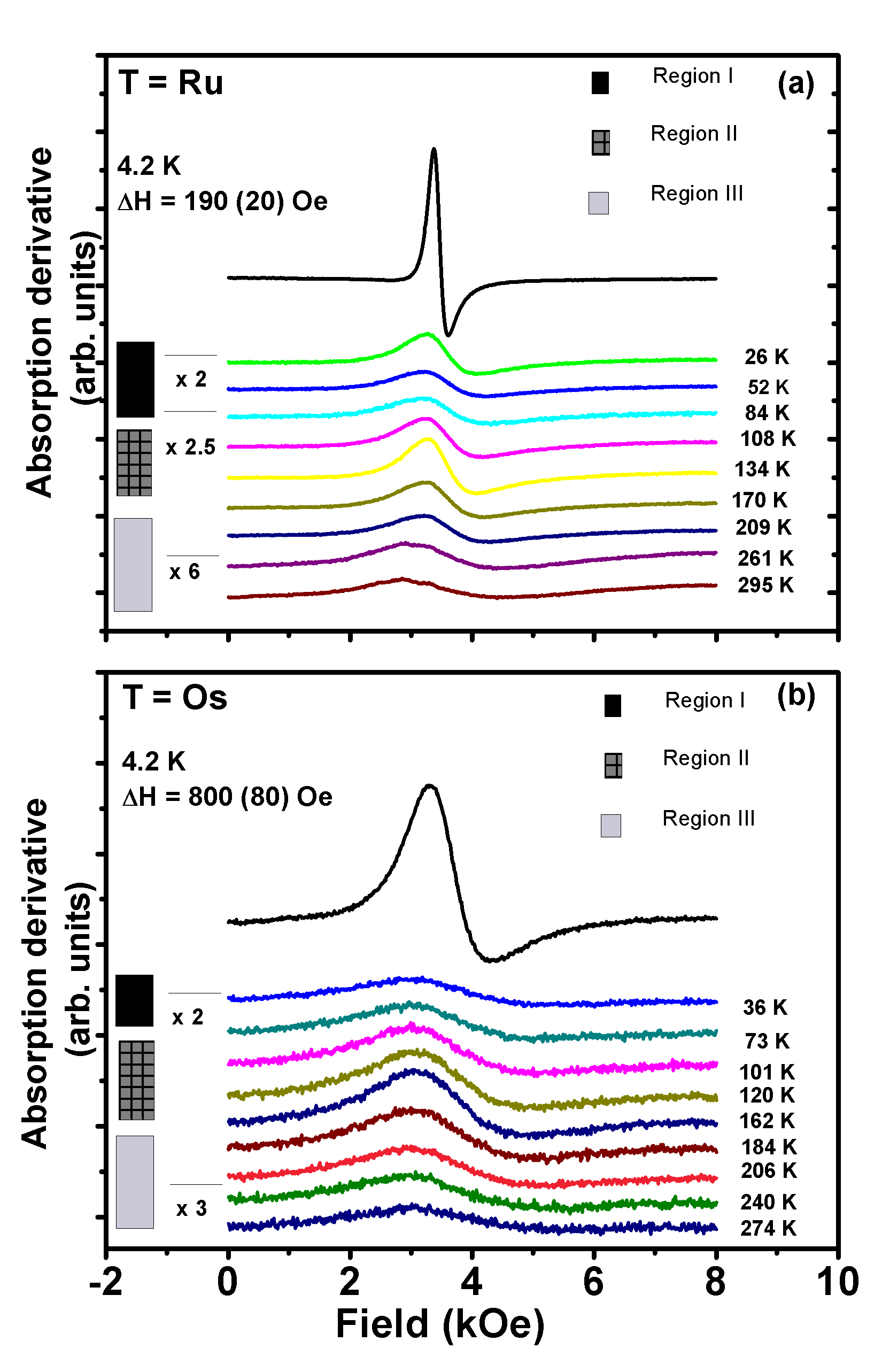} 
\par\end{centering}

\caption{X- band ESR spectra of EuT$_{4}$Sb$_{12}$ (T $=$ Ru, Os). Figures
\ref{fig:ESRspectraSku}(a)-(b) give an overview of the spectra evolution
in the temperature interval $4.2\lesssim T\lesssim300$ K and show
in detail the spectrum for $T=4.2$ K. As the temperature increases,
the spectra broaden (Region I) and then undergo a narrowing process
(Region II), reaching temperatures (Region III) where a Korringa like
relaxation process describes the behavior of $\Delta H$ as $T$ increases
further. 
We also indicate the magnifying factor for each region, in the lower
left corner of the figures.\label{fig:ESRspectraSku} }

\end{figure}

\begin{figure}
\begin{centering}
\includegraphics[scale=0.22]{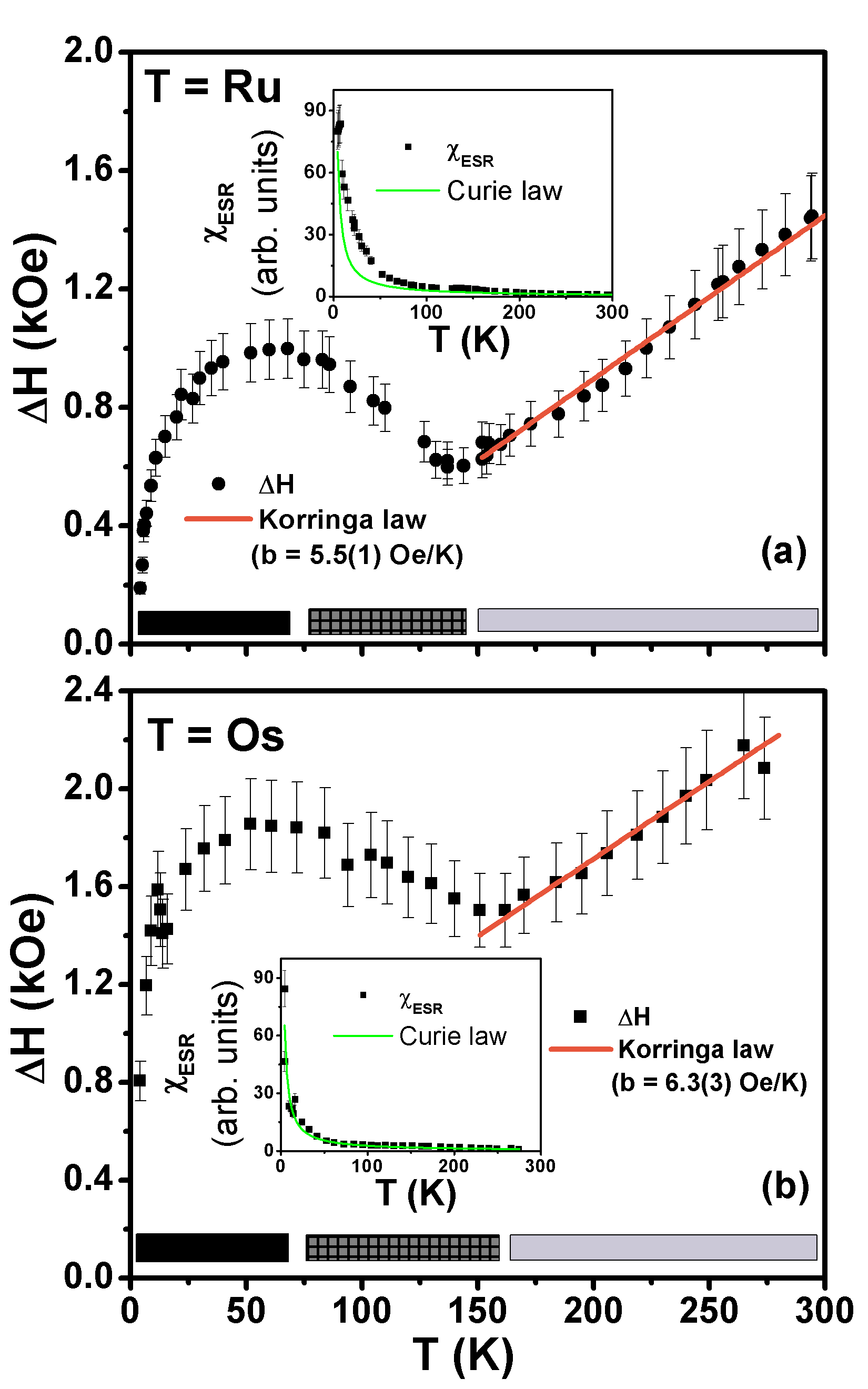} 
\par\end{centering}

\caption{X- band ESR linewidth of EuT$_{4}$Sb$_{12}$ (T $=$ Ru, Os). The
same regions I, II and III discussed in fig \ref{fig:ESRspectraSku}(a)-(b)
are indicated in the above figures. These data show clearly the existence
of such distinct regions. In the insets of figures \ref{fig:ESRlwSku}(a)-(b),
we present the X- band ESR intensity ($\chi_{ESR}$ ). It is shown
that in all regions, $\chi_{ESR}$ closely follows a Curie like behavior,
which indicates that our resonance is due to localized electronic
states. \label{fig:ESRlwSku} }

\end{figure}

\begin{figure}
\begin{centering}
\includegraphics[scale=0.22]{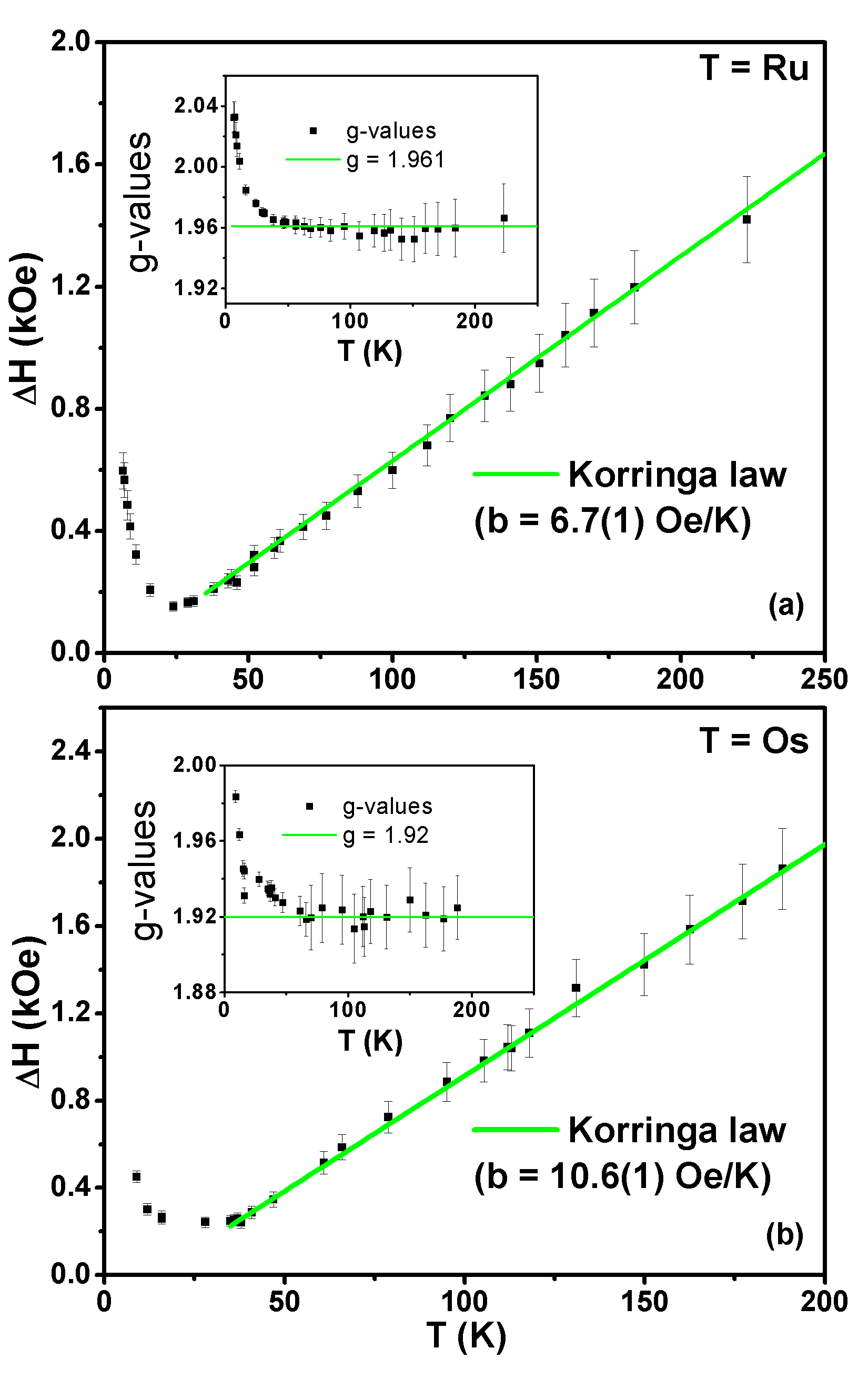} 
\par\end{centering}

\caption{Q- band ESR linewidth of EuT$_{4}$Sb$_{12}$ (T $=$ Ru, Os). Here,
no clear relation with the rattling modes is observed. For T$\gtrsim30$
K, the spectra evolve as in a Korringa-like relaxation and the slowing
down of the relaxation is seem below this temperature. The onset of
the later phenomenon starts at a slightly higher temperature in the
case T$=$Os. . In the insets of figures \ref{fig:ESRSkuQband}(a)-(b),
we present the Q-band ESR g-values. It is shown that at $T\approx35$
K and $T\approx50$, for T $=$Ru and T$=$Os, respectively, there
is an increase of the g-values reflecting the development of a ferromagnetic
internal field. \label{fig:ESRSkuQband} }

\end{figure}

\begin{figure}
\begin{centering}
\includegraphics[scale=0.22]{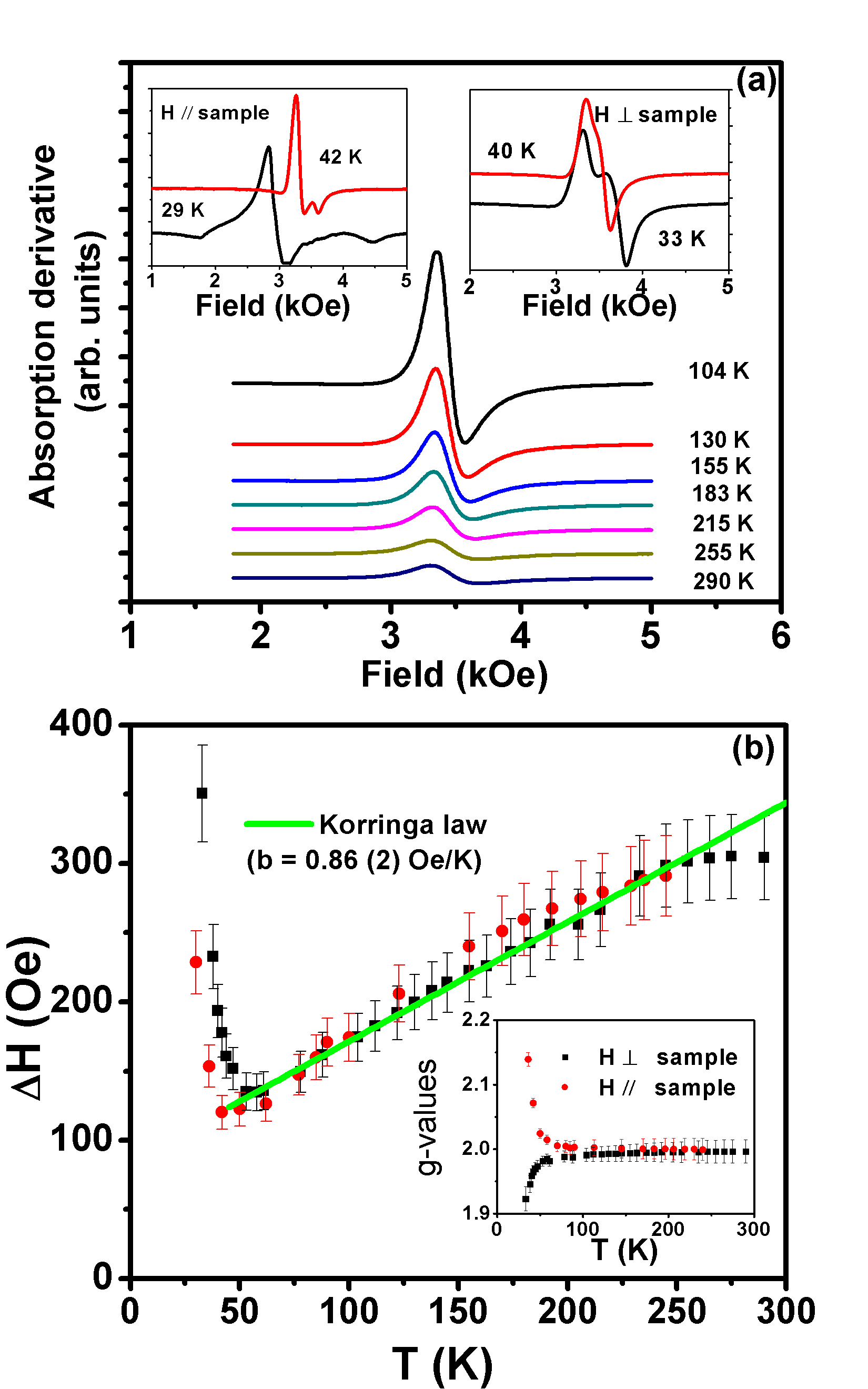} 
\par\end{centering}

\caption{X- band ESR measurements of $\beta$-Eu$_{8}$Ga$_{16}$Ge$_{30}$.
Figure \ref{fig:ESRclath}(a) gives an overview of the spectra in
the investigated temperature interval. At high temperatures, the spectra
is isotropic and the ESR intensity (not shown) follows a Curie law.
As the magnetic transition is approached, strong anisotropy develops,
as demonstrated by the two insets. Figure \ref{fig:ESRclath} (b),
show that above $T=50$ K, a Korringa like relaxation dominates the
behavior of $\Delta H$, whereas at low $T$, $\Delta H$ broadens
due to magnetic fluctuations. The inset shows strong anisotropy reflected
in the $g$-values. \label{fig:ESRclath} }

\end{figure}

Figures \ref{fig:ESRspectraSku}, \ref{fig:ESRlwSku} and \ref{fig:ESRSkuQband}
present the relevant experimental results for the skutterudite compounds
and figure \ref{fig:ESRclath} gives a brief account of our results
for the clathrate compound. In figures \ref{fig:ESRspectraSku}(a)
and \ref{fig:ESRspectraSku}(b), we give an overview of the X-band
ESR spectra obtained for EuT$_{4}$Sb$_{12}$ (T $=$ Ru, Os). It
is well seen that the thermal activation induces a fast broadening
of the spectra until $T\approx75$ K, when a narrowing process sets
in up to $T\approx150$ K. Above this $T$, the spectra broaden as
in a Korringa relaxation \citep{Whitebook, Barnes}.

As pointed out in these figures, there are 3 regions inside which
the spectra evolve with distinct characteristics. The behavior in
region III ($T\gtrsim150$ K) is expected to occur in any metal, but
the behavior found in regions I and II is unique. For T $=$ Fe, only
the ferromagnetic modes (for $T\lesssim T_{c}=90$ K) were observed
for this compound (not shown). The non-observation of the ESR signal
in EuFe$_{4}$Sb$_{12}$ may be related to one or both of the following
reasons: i) as presented in figures \ref{fig:ESRlwSku}(a) and \ref{fig:ESRlwSku}(b)
(and also in \ref{fig:ESRSkuQband}(a) and \ref{fig:ESRSkuQband}(b))
$\Delta H$ found in these compounds roughly scale with $T_{C}$ ($\Delta H_{Ru}/\Delta H_{Os}\approx T_{C}^{Ru}/T_{C}^{Os}$).
In a extrapolation, this would mean that $\Delta H_{Fe}$ would be
of the order of several kOe, that would prevent its observation; ii)
EuFe$_{4}$Sb$_{12}$ is a ferrimagnetic compound \citep{Krishnamurthy}
where the ordered Fe 3$d$ orbitals partially compensates the Eu$^{2+}$
moments. The effect of this local antiferromagnetic coupling between
the Fe 3$d$ moments and the Eu$^{2+}$ 4$f$ moment may be a huge
shift of the resonance that would also preclude its observation. 

Figures \ref{fig:ESRlwSku}(a) and \ref{fig:ESRlwSku}(b) give a better
view of the above cited regions by presenting the X-band ESR $\Delta H$
for both compounds. The behavior of $\Delta H$ resembles a coherence
peak, as found in some NMR experiments \citep{Whitebook}, when a
phase transition is approached. From region I, for both compounds,
as $T$ increases, $\Delta H$ broadens by $\approx800$ Oe. Following
this broadening, around $T\approx75$ K, both spectra gradually narrow
by $\approx300$ Oe (region II). At $T\approx150$ K, they broaden
in a Korringa like relaxation as evidenced by the fitting curve (region
III). The measured Korringa rates are $b=5.5(1)$ Oe/K for T $=$
Ru, and $b=6.3(3)$ Oe/K for T $=$ Os.

The insets in these figures show the normalized ESR intensity ($\chi_{ESR}(T)/\chi(T=300$
K$)$ for both compounds . A comparison with the Curie law shows that
in all regions the spectra behave as a resonance arising from localized
electronic states. This comparison also suggests that a small enhancement
of $\chi_{ESR}(T)$ is in order at low temperatures. This could be
ascribed to the proximity of ferromagnetic transition. The ESR intensity
was also normalized taking into account the variation of the skin
depth resulting from the $T$-dependency of resistivity \citep{Bauer}.
In the experiment with the powdered sample (EuRu$_{4}$Sb$_{12}$),
where the skin depth is bigger than the size of grains, $\chi_{ESR}(T)$
also follows a Curie-like behaviour. 

The similarities between the broadening and narrowing phenomena in
regions I and II in both compounds indicate that these phenomena should
have a common origin. In these compounds, the filler Eu atom is known
to rattle with Einstein temperatures $\theta_{E}=78$ K (T $=$ Ru)
and $\theta_{E}=74$ K (T $=$ Os) \citep{Bauer, Cao}. These temperatures
are very similar to those where $\Delta H$ approaches its maximum
before starting to narrow. We suggest that the thermal activation
of the rattling motion gives rise to the behavior of $\Delta H$ in
regions I and II.

The thermal activation has its origin in the energy dispersion of
an Einstein (harmonic) oscillator $E=(n+\frac{1}{2})\hbar\omega_{E}$
($\omega_{E}=\frac{k_{B}\theta_{E}}{\hbar}$) which, together with
the proper statistical distribution, gives the number of active oscillators
at a given temperature. Hence, in region I, there are rattling and
non-rattling resonating ions. At low-$T$, the lines are relatively
narrower due to the spin-spin Exchange interaction \citep{Abragan}.
As $T$ increases, the inhomogeneity of the spin dynamics, implied
by the presence of rattling and non-rattling resonating centers, leads
to the broadening of $\Delta H$. In region II, most of the ions are
rattling, and an homogenization process ensues, leading to narrowing
of $\Delta H$. In region III, virtually all ions are rattling and
the system has become quite homogeneous.

As a plausible scenario for the origin of this correlation between
the rattling modes and $\Delta H$, we referred to the Ruderman-Kittel-Kasuya-Yosida
(RKKY) interaction, which gives the exchange coupling between the
localized spins in rare earth metals \citep{Whitebook}. In skutterudites,
the localized spins oscillate in their Einstein frequencies and this
oscillation may give rise to random anisotropies in the exchange coupling
$J$ of the RKKY interaction, resulting in an inhomogeneous spin dynamics
of the Eu$^{2+}$ spins. Hence, in region I, the anisotropies of $J$
broaden the resonance lines and prevent the narrowing of the spectra
to occur. Once the rattling is fully activated, a narrowing process
takes place, giving rise to the behavior of $\Delta H$ in region
II. In region III, the narrowed spectra relax as in a Korringa-like
relaxation. 

The above discussed behavior of the system is very similar to a spin-glass,
as probed by an ESR experiment \citep{Dahlberg} with a glass transition
temperature $T_{G}^{*}\approx\theta_{E}$. For temperatures well below
$\theta_{E}$, the rattling is not activated, and the low-$T$ $\Delta H$
is only due to dipolar interactions and the spin-spin exchange narrowing.
However, with increasing temperature, the spin-glass behavior manifests
in the evolution of $\Delta H$. In this sense, the peaks of $\Delta H$,
as shown in Fig. \ref{fig:ESRlwSku}, are indeed coherence peaks,
although we cannot claim that they express a phase transition to a
true spin-glass state.

In contrast to the above described unique behavior of the X-band spectra,
the Q-band measurements, figures \ref{fig:ESRSkuQband}(a) and \ref{fig:ESRSkuQband}(b),
do not present this conspicuous influence of the Eu$^{2+}$ rattling
behavior. For $T\gtrsim45$ K, the spectra have a $\Delta H\approx250$
Oe and evolves as in a Korringa-like relaxation with $b=6.7(1)$ Oe/K
and $b=10.6(1)$ Oe/K for T $=$ Ru and T $=$ Os, respectively. In
this $T$- region, the relatively narrow $\Delta H$ allows for a
precise determination of the $g$-values. The results are $g=1.961(3)$,
for T $=$ Ru, and $g=1.92(1)$, for T $=$Os. In addition, at low-$T$,
both values increase, reflecting the internal fields of the low-$T$
ferromagnetic state previously reported for the compounds \citep{Bauer}.
These g-values reflect a huge g-shift ($\Delta g=g_{insulator}-g_{experiment}$),
which is due to the Exchange interaction with conduction electrons
(a polarization effect analogous to the Knight shift in NMR experiments),
when compared to $g=1.993(2)$ found in insulators \citep{Barnes}.
This negative $\Delta g$ is direct evidence for a covalent mixing
between the Eu$^{2+}$ $f$ orbitals and the conduction band $d$
orbitals. 

In a single band picture \citep{Barnes}, these $\Delta g$ imply
a Korringa rate of $b\approx24$ Oe/K and $b\approx124$ Oe/K for
T $=$ Ru and T $=$ Os, respectively. This strongly contrasts with
the experimental results, reflecting the existence of a $q$-dependency
of the exchange coupling $J(q)$ \citep{Taylor}. This is evidence
that the Eu-Eu interaction is not due to a constant $J$ exchange
coupling, giving support to our claim of an inhomogeneous RKKY interaction. 

An alternative scenario for a correlation of the rattling behavior
with the ESR spectra would be related to crystal field inhomogeneities.
In this picture, the Eu$^{2+}$ while performing its excursion within
the cage would experience slightly different crystal field parameters
and/or a crystal field environment of lower symmetry. This was shown
to be the case in our previous work on the Yb$^{3+}$ resonance \citep{GarciaRat}.
In this work, however, the broadening of $\Delta H$ in regions I
and II, instead of being enhanced, is suppressed at higher fields.
It appears that the rattling frequencies are too high, in comparison
with the ESR frequencies, and the spatial inhomogeneities are completely
averaged out. 

We ascribe the lack of clear signatures of the rattling behavior in
the Q-band ESR spectra to the relatively high fields ($H\approx1$
T) used in the experiment. As observed in ESR experiments in a true
spin-glass system \citep{Dahlberg}, higher fields tend to quench
the random anisotropy contribution. Our claim is that a field as high
as $1$T is enough to suppress the small anisotropies implied by the
Eu$^{2+}$ dynamical behavior thus suppressing the broadening of the
spectra when $T_{G}^{*}$ is approached. 

Figure \ref{fig:ESRclath} shows that no comparable phenomenon takes
place in the clathrate compound. Figure \ref{fig:ESRclath}(a) shows
that at low-$T$, the spectra are anisotropic due to the proximity
of the magnetic transition ($T_{C}=35$ K \citep{Sales2}). Above
this low-$T$ region (T $\gtrsim70$ K), the spectra are isotropic
with $g=1.999(3)$ and broaden in a Korringa like relaxation. Figure
\ref{fig:ESRclath}(b) evidences the relatively narrow ESR line when
compared to the skutterudite compounds. As the transition $T$ is
approached, and below this temperature, $\Delta H$ broadens rapidly
and continuously. The $g$-values also reflect clearly the low-$T$
anisotropy and magnetic transition.

No clear connection with the Eu$^{2+}$ rattling modes is observed
in this experiment. Given the specific features of the behavior of
the filler atom in this clathrate (\emph{off-center} rattling), we
would expect more signatures arising from crystal field inhomogeneities
in this experiment than in the experiments with the skutterudites.
This, and the Q-band measurements, lead us to rule out crystal field
effects as the origin of the phenomena observed in figures \ref{fig:ESRspectraSku}(a)-(b)
and \ref{fig:ESRlwSku}(a)-(b). Both results considered together,
give strength for our claim that the unique behavior of the ESR spectra
of the skutterudites are due to the coupling between the rattling
modes and electronic degrees of freedom. 

Although we observe a metallic relaxation, the process is slow ($b=0.86$
Oe/K), which agrees with previous results pointing that Eu$_{8}$Ga$_{16}$Ge$_{30}$
is a poor metallic system \citep{Sales2}. This Korringa rate is compatible
with the measured $\Delta g=0.006(5)$, pointing to a $q$-independent
$J_{fs}$ constant and absence of multi-band effects. 

It is noteworthy that $\Delta H$ in the skutterudites and in the
clathrates are of the same order of magnitude at low-$T$ (see figures
\ref{fig:ESRclath} and \ref{fig:ESRSkuQband} around $T\approx50$
K). In general, $\Delta H$ in clean concentrated metallic systems
will be determined by strong dipolar interactions, which broaden the
line, and spin-spin Exchange narrowing effects. Since the Eu$^{2+}$
ions are slightly more apart from each other in the clathrates, one
would expect dipolar interaction to be weaker in the clathrates and
hence $\Delta H$ would be smaller, as it is verified. Since the latter
two effects are temperature independent, the significant difference
in $\Delta H$ at high-$T$ should be ascribed to the different Korringa
rates of the Korringa-like relaxation \citep{Barnes}, discussed above.
The exact nature of a relaxation process in a concentrated metallic
system is very hard to determine. However, it should be related to
an Exchange interaction with conduction electrons, and that is why
we are carefull to refer to the linear broadening of $\Delta H$ as
a {}``Korringa-like'' process. All the reported material parameters
relevant to this coupling, such as the density of electronic states
at the Fermi surface \citep{Bauer,Sales2}, favors the idea that this
coupling (and therefore the relaxation) should be stronger in the
skutterudites.

Contrary to the skutterudites studied here, we note that for this
compound, Raman scattering studies\citep{Takasu} have shown that
the Eu$^{2+}$ rattling energy ($\theta_{E}\approx25$ K) is lower
than that of the magnetic ordering ($T_{C}=36$ K). Hence, a spin-glass
like behavior below $\theta_{E}$ may also be prevented from occurring
by the magnetically ordered state. Furthermore, the itinerant $d$
orbital contribution to the Fermi surface, in the case of the skutterudites,
is most likely originated from the Ru (Os) $d$ atomic orbitals, indicating
that the skutterudite cage is somehow stiffer (more correlated) than
the clathrate cage. 
Indeed, the $\Delta g$ analysis point for a greater coupling of localized
and itinerant states in the skutterudites, which should be important
to correlate the rattling behavior with the Eu-Eu interaction.

In some skutterudites, there are earlier reports for temperature dependence
of the valence state of the guest ion, as was reported for the related
compound YbFe$_{4}$Sb$_{12}$ \citep{Dilley}. This result raises
the question about the valence state of the Eu ion on the compounds
here investigated. We should point out, however, that the issue of
the Yb valence state in YbFe$_{4}$Sb$_{12}$ was subsequently revised
in the literature \citep{Schenelle3}, determining that the Yb ions
are in a stable divalent state. In our compounds, no significant fraction
of Eu$^{3+}$ was found for EuT$_{4}$Sb$_{12}$ (T $=$Ru, Os) nor
it was found any sign for temperature dependency of the Eu$^{2+}$
fraction in these materials \citep{Bauer}. In EuFe$_{4}$Sb$_{12}$,
a fraction of $10-15$ \% of Eu$^{3+}$ was reported. However, again,
no sign for temperature dependency of this fraction was found, even
in detailed studies of X-ray absorption spectroscopy \citep{Krishnamurthy, Krishnamurthy2}.
The insets of figures \ref{fig:ESRlwSku}(a)-(b) show a Curie-like
behavior for the ESR intensity, that should also be taken as another
piece of evidence for a nearly stable Eu$^{2+}$ configuration.

\section{Conclusion}


In conclusion, we have provided evidence, from Eu$^{2+}$ X-band ESR
measurements, for inhomogeneous spin dynamics of the Eu$^{2+}$ spins
in the EuT$_{4}$Sb$_{12}$ (T$=$ Ru, Os) skutterudites, triggered
by the Eu$^{2+}$ rattling modes. Our Q-band measurements were discussed
in terms of presenting evidence for $q$-dependence of $J(q)$ exchange
coupling and for covalent mixing between localized $f$ and itinerant
$d$ orbital states. These two findings were related, respectively,
to the inhomogeneous spin dynamics of the Eu$^{2+}$ spins and as
an evidence for a stronger coupling between localized and itinerant
states in the skutterudites than in the clathrate. 

Our picture is that the behavior of $\Delta H$ in the X-band ESR
measurements is due to an electron-phonon coupling between the rattling
phonon modes and the electronic degrees of freedom. In regions I and
II of figures \ref{fig:ESRlwSku}(a) and \ref{fig:ESRlwSku}(b), the
itinerant $d$ orbital electronic states are coupled with oscillating
and non-oscillating 4$f$ electronic states, which gives rise to random
anisotropies of the exchange coupling and subsequently to the inhomogeneous
spin dynamics found in these regions. 

In analogy with the phonon-glass scenario discussed for cage systems,
we discussed our findings in the framework of a spin-glass type of
spin dynamics. In this sense, high fields tend to suppress the ESR
line broadening when $T_{G}$ is approached, by quenching the random
anisotropies of these systems. We suggested that a similar effect
takes place in our experiment, preventing the observation of inhomogeneous
dynamics of the Eu$^{2+}$ spins in Q-band. Further theoretical and
experimental investigation may unravel the exact characteristics of
this spin-glass like state and also explore the perspective of other
phenomena emerging from the interaction between conduction electrons
and the rattling phonon modes.

\begin{acknowledgments}
The authors would like to thank the Brazilian agencies FAPESP and
CNPq, for the financial support, and Brian C. Sales, for valuable
discussions on the sample growth methods. C. Rettori also acknowledge
CAPES for financial support. 
\end{acknowledgments}

\end{document}